\documentclass[fleqn,usenatbib]{mnras}

\usepackage{newtxtext,newtxmath}

\usepackage[T1]{fontenc}
\usepackage{tablefootnote}
\usepackage{adjustbox}

\usepackage{graphicx}	
\usepackage{amssymb}	
\usepackage{multirow} 
\usepackage{color}

\newcommand{\coo}{CO$_2$}
\newcommand{\oo}{O$_2$}

\title[Thermal desorption from forsterite(010)]{Thermal desorption of astrophysically relevant molecules from forsterite(010)}

\author[T. Suhasaria et al.]{
T. Suhasaria,
J. D. Thrower,
H. Zacharias \thanks{E-mail: john.thrower@uni-muenster.de (JDT), H.Zacharias@uni-muenster.de (HZ)}
\\
Physikalisches Institut, Westf{\"a}lische Wilhelms-Universit{\"a}t, 48149 M{\"u}nster, Germany
}

\date{}

\pubyear{2017}

\begin{document}
\label{firstpage}
\pagerange{\pageref{firstpage}--\pageref{lastpage}}
\maketitle

\begin{abstract}
We present temperature programmed desorption (TPD) measurements of CO, CH$_4$, O$_2$ and CO$_2$ from the forsterite(010) surface in the sub-monolayer and multilayer coverage regimes. In the case of CO, CH$_4$ and O$_2$, multilayer growth begins prior to saturation of the monolayer peak, resulting in two clearly distinguishable desorption peaks. On the other hand a single peak for CO$_2$ is observed which shifts from high temperature at low coverage to low temperature at high coverages, sharpening upon multilayer formation. The leading edges are aligned for all the molecules in the multilayer coverage regime indicating zero order desorption. We have extracted multilayer desorption energies for these molecules using an Arrhenius analysis. For sub-monolayer coverages, we observe an extended desorption tail to higher temperature.  Inversion analysis has been used to extract the coverage dependent desorption energies in the sub-monolayer coverage regime, from which we obtain the desorption energy distribution. We found that owing to the presence of multiple adsorption energy sites on the crystalline surface the typical desorption energies of these small molecules are significantly larger than obtained in previous measurements for several other substrates. Therefore molecules bound to crystalline silicate surfaces may remain locked in the solid state for a longer period of time before desorption into the gas phase.
\end{abstract}



\section{Introduction}

A detailed understanding of molecular interactions on dust grain surfaces is of prime importance in a variety of astrophysical environments. In cold, dense regions of the interstellar medium (ISM) and on cometary bodies, molecules are frozen in the solid state and form icy mantles around dust grains. In addition to the carbonaceous grain population, silicate minerals of the olivine family are thought to be a common constituent of such dust grains. Cometary silicates are an unequilibrated and heterogenous mixture of crystalline and amorphous silicates \citep{hanner99}. In cold, dense regions of the ISM, silicates are generally thought to be amorphous \citep{li02}. Crystalline silicates are believed to be formed by high temperature annealing and predominantly reside close to evolved stars \citep{harker02, gail04}. \citet{kemper04} suggest a value of only 2.2\% of crystalline silicates which must have been injected into the ISM by the stellar winds after their formation. However, based on recent observations, crystalline silicates have been detected as interstellar grains by the Stardust Mission \citep{westphal14} and also in the ISM along several lines of sight towards the Galactic Centre \citep{wright16}. These authors showed further evidence for the presence of a measurable quantity of crystalline silicates in the envelopes of young stellar objects (YSOs) as well as along diffuse and dense cloud sight lines. They suggested that the quantity of crystalline silicates in ISM might be underestimated as their direct observation is complicated as a result of spectral overlap with other solid state species\citep{wright16}. Forsterite (Mg$_2$SiO$_4$) is a convenient analogue for magnesium-rich crystalline silicates, serving as a model surface with which to investigate the interactions between weakly bound physisorbed molecules and interstellar grain surfaces.

During the evolution of interstellar clouds, thermal processing of the icy mantles on grain surfaces commonly occurs, as is also the case for cometary ices when comets move closer to the sun \citep{wyckoff82}. The timescale for the return of ice species to the gas phase through thermal desorption depends on the strength of the interaction between adsorbed molecules and the grain surface as well as, for multilayer ices, the interactions between adsorbate molecules. Thus, measurements of the adsorption energy $E_{\mathrm{ads}}$ provide valuable input for astrochemical models of such environments. Temperature programmed desorption (TPD) measurements provide a means to extract the kinetic desorption parameters for molecules adsorbed on model grain surfaces. Several independent studies in recent years \citep{katz99, thrower09, noble12, clemens13, smith14b, suhasaria15, smith16, he16} have demonstrated that, in addition to the chemical nature of the surface, surface morphology plays a key role in determining the kinetic parameters for desorption. For example, coverage dependent kinetics arise where a grain surface displays multiple adsorption sites with differing adsorption energies.

In addition to water ice, CO, CO$_2$, CH$_4$ and O$_2$ are also important components of interstellar and cometary ices \citep{mumma11}. CO is the second most abundant molecule after H$_2$O and has a relative abundance of up to 85\% in low mass YSOs and up to 30\% in comets \citep{boogert15}. Solid CO$_2$ typically shows an abundance of 12-50\% with respect to water in low mass YSOs and 4-30\% in comets \citep{boogert15}. CH$_4$ is the simplest organic molecule and its solid phase abundance with respect to H$_2$O ice is found to be 1-11\% in low mass YSOs and between 0.4 and 1.6\% in comets \citep{boogert15}. O$_2$ is expected to be an abundant molecule in the ISM \citep{larsson07}, although direct observation is difficult as it lacks a permanent dipole moment. Nevertheless, O$_2$ is considered to be involved in the formation of H$_2$O on grain surfaces \citep{goldsmith11}. Recently, the cometary sampling and composition (COSAC) experiment onboard the Philae lander of the Rosetta Mission has confirmed the presence of CO, CO$_2$ and CH$_4$ along with H$_2$O and other organic species in cometary ices \citep{goesmann15}. 

In recent years the desorption kinetics of these simple species has been studied experimentally for carbonaceous \citep{ulbricht06, burke10, edridge13, smith16}, water ice \citep{collings03, noble12, edridge13, smith16, he16}, amorphous silicate \citep{noble12} and amorphous silica \citep{collings15} surfaces. To date the desorption of these species on crystalline silicate surfaces has not been investigated experimentally, except for one study of CO$_2$ on the olivine(011) \citep{smith14b}. The adsorption of CO \citep{escamilla17} and CO$_2$ \citep{kerisit13} on forsterite surfaces has, however, been investigated using density functional theory (DFT). Recently, we have experimentally studied the desorption kinetics of NH$_3$ from two forsterite(010) surfaces prepared by cleaving and cutting the surface, the latter resulting in a surface with a significantly higher degree of surface heterogeneity. In both cases, a large distribution of desorption energies was observed, leading to a significantly longer desorption timescale compared to that suggested by a single desorption energy. Here we have extended this work to other, more weakly bound astrophysically relevant molecules that desorb at much lower temperatures from the cleaved forsterite(010) surface.

\section{Experimental methods}

Experiments were performed in an ultra-high vacuum chamber with a base pressure of 5$\times10^{-10}$ mbar which is described in detail elsewhere \citep{suhasaria15}. The sample is clamped to the end of a liquid helium cryostat which allows cooling to $<15$ K. The sample is locally heated using a 50 W halogen bulb situated just behind the sample holder. The temperature of the sample is determined with a precision of better than 0.5 K by a calibrated silicon-diode which is mounted close to the sample on the sample holder. The sample is heated with the aid of proportional-integral-derivative control software written in LabView.  A quadrupole mass spectrometer (QMS; Hiden Analytical HAL 7) detects the molecules desorbing from the sample surface. The QMS is contained within a glass shield with an entrance aperture of 2 mm which reduces the detection of molecules desorbing from surfaces other than that of the sample of interest.

\begin{figure*}
\includegraphics[width=\textwidth]{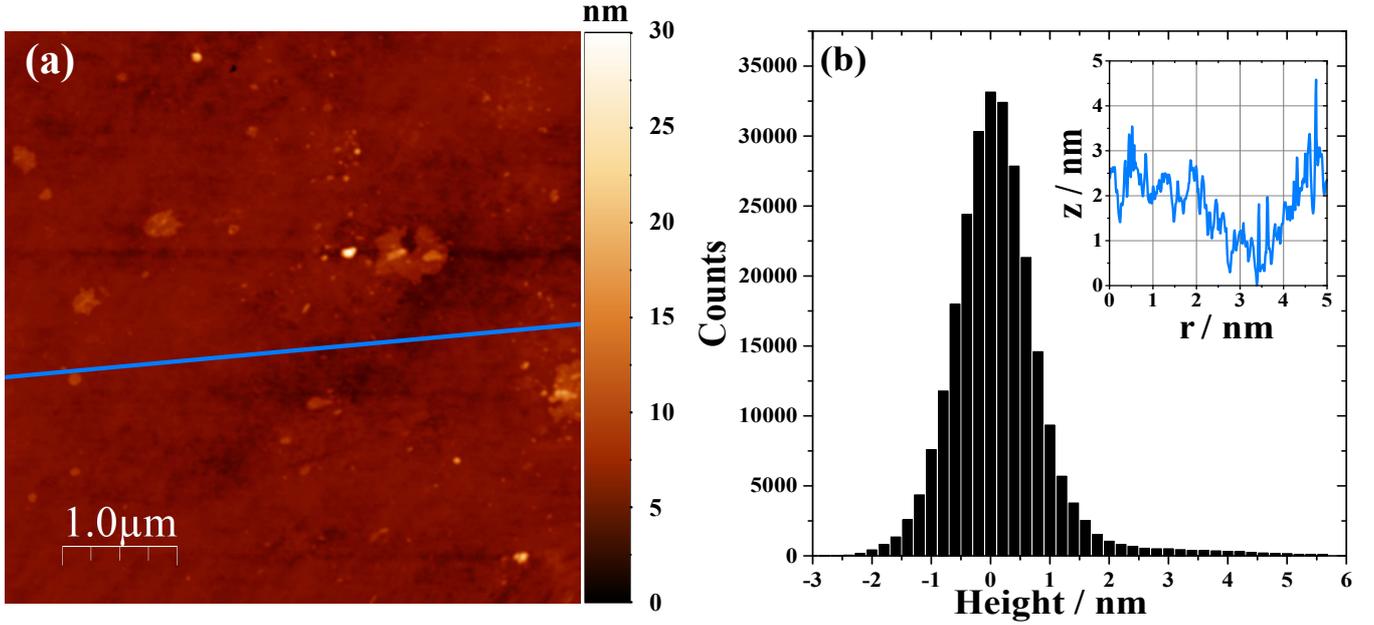}
\caption{ (a) Tapping mode topographical AFM image of $5\times5 \mathrm{\mu{m}}^2$ region of the forsterite(010) surface. (b) Height distribution with respect to the flat terrace. The inset shows the profile taken along the blue line in (a), represented by $r$.}
\label{fig1}
\end{figure*}

The sample was a synthetic forsterite single crystal, cleaved to expose the most stable forsterite(010), as confirmed through reflective diffraction measurements \citep{suhasaria15,king10}. To understand the surface morphology, tapping mode atomic force microscopy (AFM) was used. The AFM images obtained were visualized using the WsxM 5.0 software package \citep{horcas07}. Figure \ref{fig1} (a) shows a topographical 5$\times5 \mathrm{\mu{m}^2}$ region of the cleaved forsterite(010) surface. The image indicates that the cleaving of forsterite surface yields atomically flat terraces with some atomic defects. Figure \ref{fig1} (b) shows the height distribution, centred around the most common height, which demonstrates a maximum height variation of \textit{ca.} $\pm 2$ nm. This is also clear in the line scan along the blue line displayed in the inset to Figure \ref{fig1} (b). The root mean square (rms) roughness across the surface is found to be 0.8 nm. In comparison, the cut forsterite(010) surface used in our previous study \citep{suhasaria15} has a rms roughness of 81 nm and a mean height variation of 326.1 nm. Thus, the cleaved surface is significantly smoother.

Prior to performing thermal desorption measurements, the sample was heated to 465 K in UHV for several minutes to remove surface contaminants, and to degas the heating filament. The sample was then rapidly cooled to \textit{ca.} 17 K prior to exposure to the adsorbate of interest. A small UHV dosing chamber is connected to the main chamber and is isolated by a gate valve. The pressure in the dosing chamber is measured with a calibrated spinning rotor gauge. Gases from the dosing chamber are introduced into the main chamber by means of long tube terminated with a 50 $\mathrm{\mu}$ m aperture. The sample is brought to within 5 mm of the dosing tube to ensure that the gas is dosed directly on the sample, strongly reducing adsorption on other surfaces of the sample holder as confirmed by only a small pressure rise to \textit{ca}. 5.5$\times10^{-10}$ mbar in the main chamber during dosing. We note that dosing from a small aperture in this way will lead to some degree of non-uniformity in the film growth rate across the surface. Assuming a cosine flux distribution, we estimate that the flux at the surface varies by less than 10 \% over \textit{ca.} 85 \% of the surface. Any effects of this non-uniformity are minimized by the aperture to the QMS which allows us to constrain the detection to molecules desorbing from the uniform region of the film. For each molecule the same initial pressure in the dosing chamber was used, with the exposure being determined by the duration of dose. The gases, carbon monoxide (CO) (99.97\%, Westfalen AG), oxygen (O$_2$) (99.999\%; Westfalen AG), methane CH$_4$ (99.995\%; Westfalen AG) and carbon dioxide (CO$_2$) (99.995\%; Westfalen AG), were used directly without further purification.

\section{Results and analysis}

\subsection{Coverage determination}\label{coverage_determination}

In order to describe the desorption kinetics it is necessary to determine the surface coverage in terms of the number of adsorbed monolayers (ML) that result from a given exposure to the gas of interest. In order to determine the coverage for a given TPD trace we note that the mass spectrometer signal for a desorbing species is directly proportional to its rate of desorption, $r_{\mathrm{des}}$, as described by the Polanyi-Wigner equation: \citep{king75}

\begin{equation}
 r_{\mathrm{des}} = -\frac{\mathrm{d} \Theta}{\mathrm{d} t} = \nu_n \Theta^n\exp\left[-\frac {E_{\mathrm{des}}}{RT}\right]
\label{equation1}
\end{equation}
where $\nu_n$ denotes the pre-exponential factor (ML$^{n-1}$ s$^{-1}$), $n$ the desorption order, $E_{\mathrm{des}}$ the desorption energy (kJ mol$^{-1}$), $R$ the universal gas constant (kJ mol$^{-1}\rm K^{-1}$), and $T$ the surface temperature (K). $\Theta$ denotes the relative coverage of adsorbate molecules on the forsterite(010) surface in ML. We define $\Theta_\mathrm{i}$ as the initial coverage at the start of the TPD measurement that results from the adsorbate deposition, \textit{i.e.}:

\begin{equation}
\Theta_\mathrm{i} = \int{r_{\mathrm{des}}}\mathrm{d}t
\label{equation2}
\end{equation}
where $r_{\mathrm{des}}$ is in units of ML s$^{-1}$.

\begin{figure}
{\includegraphics[width=\columnwidth]{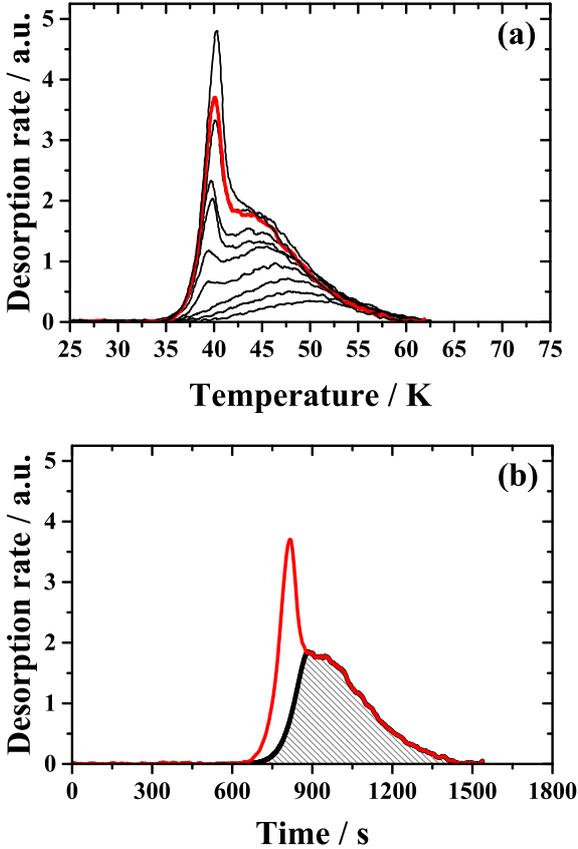}}
\caption{ (a) TPD traces for CH$_4$ desorbing from the forsterite(010) surface for increasing exposures. The TPD trace in red is used for coverage determination. (b) shows an example of a simulated first order desorption peak. The grey shaded area corresponds to the saturated monolayer, obtained by matching the simulated first order leading edge (in black) with the experimental trailing edge (in red).}
\label{fig2}
\end{figure}

Although the proportionality constant between the QMS signal and $r_{\mathrm{des}}$ is initially unknown, it can be obtained through the integration of the TPD trace corresponding to a known coverage. For example, in cases where the adsorbate first forms a monolayer on the surface prior to the growth of a multilayer film, a distinct peak in the TPD trace arises that saturates at 1 ML. However, in case of CO, CH$_4$ and O$_2$, we observe the appearance of a second desorption peak prior to that attributed to the monolayer. This indicates that multilayer growth, which results in a zero order desorption peak at lower temperature, commences prior to saturation of the monolayer peak at higher temperature. This suggests the formation of few layer islands during the film growth, rather than a simple layer by layer growth mechanism. This effect may arise to some extent as a result of the non-uniformity of the dosing flux as discussed in the Experimental Methods. Therefore, direct extraction of the integrated peak area corresponding to a coverage of 1 ML is more difficult. This is shown in Figure \ref{fig2} (a), for the desorption of CH$_4$ from forsterite(010) where the desorption peak at higher temperature corresponds to the monolayer while the other, at lower temperature, is due to desorption from CH$_4$ multilayers. The TPD trace in red corresponds to the lowest exposure that results in a saturated monolayer peak and is therefore used in the coverage determination.

In order to isolate the desorption peak associated with the saturated monolayer alone, we construct a synthetic leading edge assuming first order desorption kinetics and match this to the experimental trailing edge. We assume that the peak maximum of the monolayer peak corresponds to the desorption out of the lowest energy adsorption sites and use the Redhead peak maximum method \citep{redhead62} to extract the corresponding desorption energy. These parameters were then used to simulate the corresponding first order desorption peak using the 4th-order Runge-Kutta method to solve the Polanyi-Wigner equation. This is shown in Figure \ref{fig2} (b) where the leading edge of the simulated monolayer peak (in black) is matched to the experimental trailing edge (in red) to reconstruct the monolayer peak. Thus, the grey shaded area corresponds to an initial coverage of 1 ML and can be used, by comparison, to obtain the initial coverages for the other exposures used. The same approach was used to determine the coverages for CO and O$_2$. On the other hand, the CO$_2$ desorption shows no clear distinction between monolayer and multilayer peaks. In this case, we therefore define 1 ML according to the distinctive sharpening in the lower temperature region that arises upon the appearance of multilayer desorption with zero order kinetics. This is discussed in more detail in Section \ref{ThermalDesorption} with reference to Figure \ref{fig3}.

We note that for all four molecules, the leading edges in the TPD traces are not coincident for first few multilayers which suggests fractional order desorption process, most likely resulting from the growth of multilayer islands, further consistent with a non layer by layer growth mechanism. This effect could again arise as a result of non-uniform dosing across the sample. We cannot therefore define monolayer peak as the coverage simply at which the leading edges of the TPD curves become coincident. 

\subsection{Thermal desorption}\label{ThermalDesorption}

\begin{figure*}
\centering
    \resizebox{\hsize}{!}{\includegraphics{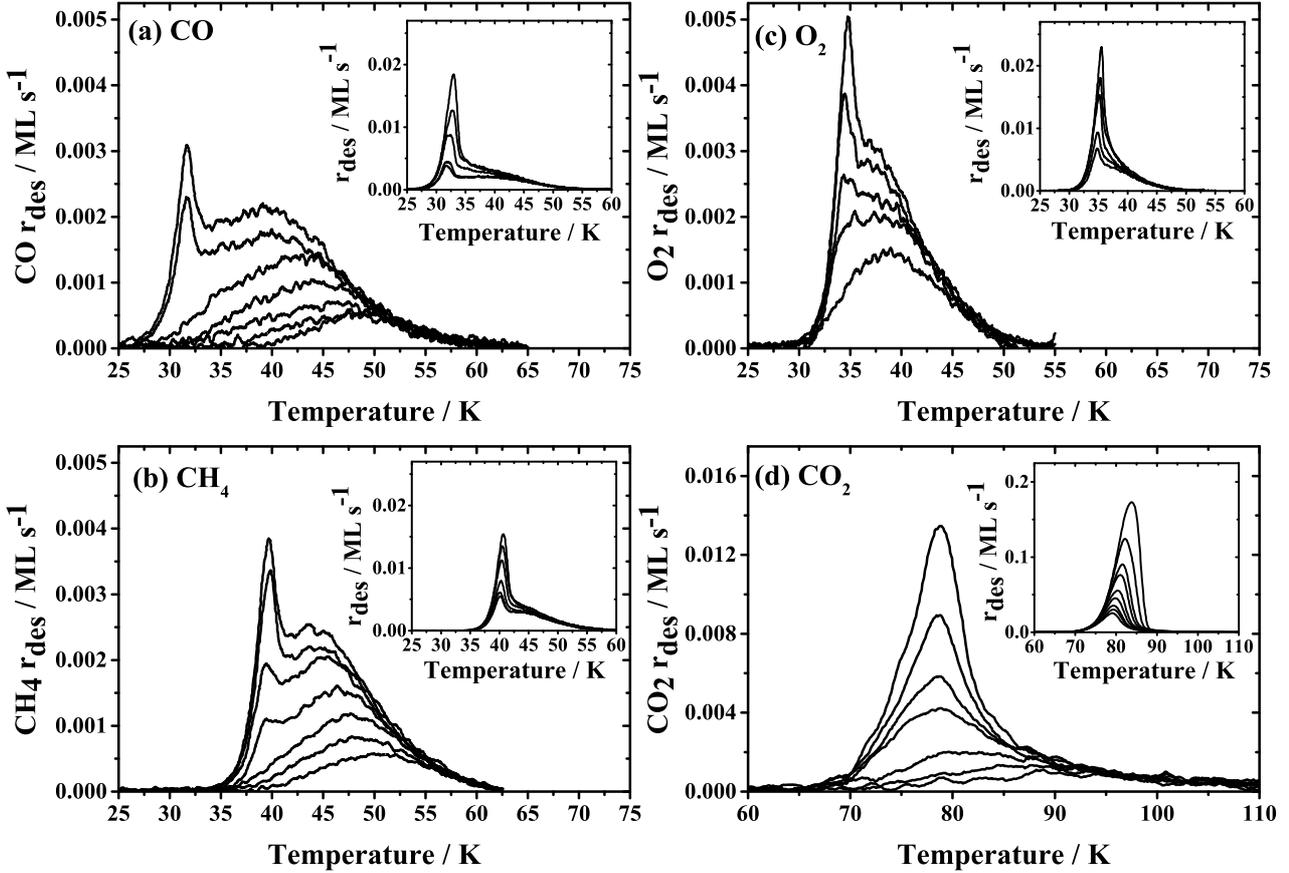}}
\caption{ (a) TPD traces for CO on forsterite(010) for increasing surface coverages of $\Theta_{\mathrm{i}}$ = 0.2, 0.3, 0.4, 0.5, 0.8, 1.0 and 1.2 ML. The inset shows the multilayer coverages of $\Theta_{\mathrm{i}}$ = 1.3, 1.4, 2.0, 2.5 and 2.8 ML.
(b) TPD traces for CH$_4$ on forsterite(010) for increasing surface coverages of $\Theta_{\mathrm{i}}$ =  0.2, 0.4, 0.5, 0.7, 0.9, 1.1 and 1.2 ML. The inset shows the multilayer coverages of $\Theta_{\mathrm{i}}$ = 1.3, 1.4, 1.5, 1.9, 2.2 and 2.3 ML.
(c) TPD traces for O$_2$ on forsterite(010) for increasing surface coverages of $\Theta_{\mathrm{i}}$ = 0.5, 0.8, 0.9, 1.0 and 1.1 ML. The inset shows the multilayer coverages of $\Theta_{\mathrm{i}}$ = 1.3, 1.7, 2.0, 2.4 and 2.6 ML.
(d) TPD traces for CO$_2$ on forsterite(010) for increasing surface coverages of $\Theta_{\mathrm{i}}$ = 0.3, 0.4, 0.6, 0.8, 0.9, 1.1 and 1.3 ML. The inset shows the multilayer coverages of $\Theta_{\mathrm{i}}$ = 1.9, 2.2, 2.5, 3.3, 3.8, 5.1, 5.9, 8.8 and 12.2 ML. All traces have been smoothed using the same adjacent averaging procedure.}
\label{fig3}
\end{figure*}

We begin by summarizing the observed thermal desorption characteristics for the four molecules investigated. Figure \ref{fig3} shows the thermal desorption traces obtained for (a) CO, (b) CH$_4$, (c) O$_2$ and (d) CO$_2$ from forsterite(010) following adsorption at 17 K. Linear heating rates of $\beta=0.03$ K s$^{-1}$ for CO, CH$_4$ and O$_2$ and $\beta=0.1$ K s$^{-1}$ for CO$_2$, respectively, were employed. In each case, traces obtained from several different exposures covering the sub-monolayer and multilayer coverage regimes are presented. The initial coverage that results from each exposure has been determined according to the procedure described in Section \ref{coverage_determination}. In all cases, the lowest exposures result in a desorption peak at high temperature that gradually extends to lower temperature with increasing exposure, leading to the development of a high temperature tail in the desorption peak. The tail extends typically about 15 K beyond the position of the peak maximum for the saturated monolayer. As this peak dominates at low exposures, and ultimately saturates with increasing exposure we attribute this peak to desorption from the first monolayer of adsorbate molecules on the forsterite(010) surface. The broad temperature range over which the desorption traces extends is reminiscent of that observed for NH$_3$ adsorbed on the same surface \citep{suhasaria15}. Such behaviour can  be attributed to either the presence of a variety of adsorption sites on the forsterite surface that give rise to a distribution of adsorption energies, or to increasing repulsive interactions between adsorbate molecules as the coverage increases. In both cases, the result is a coverage dependence of the desorption energy. Coverage dependent desorption energies have been observed previously for \textit{e.g.} CO on MgO(100) \citep{dohnalek01}, D$_2$ on amorphous solid water \citep{amiaud07}, benzene (C$_6$H$_6$) on amorphous silica \citep{thrower09}, CO$_2$ and H$_2$O on crystalline olivine(011) \citep{smith14b} and NH$_3$ on amorphous silicate \citep{he15}. In the first scenario, the shift of the desorption to lower temperature with increasing exposure suggests a sequential filling of adsorption sites from higher to lower binding energy. We note that TPD is not sensitive to the initial population of adsorption sites prior to the commencement of heating as a result of thermally activated diffusion which would result in the migration of molecules to higher energy sites during the desorption measurement. Thus, the experimental desorption profile reflects the overall distribution of binding sites on the surface. In the second scenario, the shift of the desorption to lower temperatures with increasing exposure could also be attributed to repulsive adsorbate-adsorbate interactions. These increase  as the adsorbate molecules occupy more sites on the substrate, reducing their average separation. This has been suggested previously by \citet{sallabi00} for the desorption of CO from MgO(100) by considering Monte carlo simulations and experimentally for the desorption of O$_2$ and CO on the fully oxidized TiO$_2$ (110) surface \citep{dohnalek06}.

From Figure \ref{fig3} it is clear that in the intermediate coverage regime, corresponding to the growth of the first few multilayers, the leading edges are not aligned which can be interpreted as fractional order desorption kinetics arising from island growth on the surface. We again note that this may arise as a result of the non-uniformity of dosing across the surface resulting in completion of the monolayer in some regions prior to others. As the exposure increases, all species show the formation of a sharper desorption peak at lower temperature. This peak gradually shifts to higher temperature with increasing coverage and does not saturate. These characteristics are indicative of desorption from a multilayer ice film. Furthermore, we note that this peak shows coincident leading edges, as highlighted in the insets to Figure \ref{fig3} as expected for zero order multilayer desorption. 

A clear difference is apparent in the sub-monolayer desorption of CO$_2$ compared to the other three molecules. While CO, CH$_4$ and O$_2$ show two distinct desorption peaks, CO$_2$ shows only a single peak that gradually extends to lower temperature with increasing coverage and then sharpens upon formation of the multilayer. Beyond this point the multilayer desorption peak evolves similarly to the other molecules.

Figure \ref{fig3} (a) shows the TPD traces for CO for the sub-monolayer and intermediate regimes from $\Theta_{\mathrm{i}}$ = 0.2 to 1.2 ML. The sub-monolayer peak maximum shifts from 49.3 K at $\Theta_{\mathrm{i}}$ = 0.2 ML to 40 K at $\Theta_{\mathrm{i}}$ = 1.2 ML with the multilayer peak growing in at 31.7 K prior to the saturation of the monolayer peak. The high temperature tail extends up to \textit{ca.} 65 K. The inset shows traces corresponding to  multilayer coverages of $\Theta_{\mathrm{i}}$= 1.3 to 2.8 ML. The leading edges begin to align at a total coverage of 2.0 ML. 

Figure \ref{fig3} (b) shows the TPD traces for CH$_4$ for the sub-monolayer and intermediate coverage regimes from $\Theta_{\mathrm{i}}$ = 0.2 to 1.2 ML. For sub-monolayer coverages, the peak maximum shifts from 50.5 K at $\Theta_{\mathrm{i}}$ = 0.2 ML to 44.5 K at $\Theta_{\mathrm{i}}$ = 1.2 ML, with a high temperature tail extending up to \textit{ca.} 62.5 K in all cases. The multilayer peak at 39.7 K begins to grow already at a total coverage of 0.7 ML, which again suggests the formation of islands on the surface before completion of the monolayer.  The inset shows the traces corresponding to multilayer coverages ranging from  $\Theta_{\mathrm{i}}$= 1.3 to 2.3 ML. With increasing exposure, the multilayer peak maximum shifts to higher temperatures as expected for zero order kinetics. The leading edges are aligned for coverages greater than 1.9 ML.

Figure \ref{fig3} (c) shows the TPD traces for O$_2$  for the sub-monolayer and intermediate regimes from $\Theta_{\mathrm{i}}$ = 0.5 to 1.1 ML. The sub-monolayer peak maximum shifts from 38.7 K at $\Theta_{\mathrm{i}}$ = 0.5 ML to 37.3 K at $\Theta_{\mathrm{i}}$ = 1.1 ML. The multilayer peak appears at 34.4 K for a coverage of 0.9 ML before saturation of the monolayer peak. The high temperature tail extends up to \textit{ca.} 55 K. The inset shows multilayer traces corresponding to coverages of $\Theta_{\mathrm{i}}$ = 1.3 to 2.6 ML, which again shift to higher temperature with increasing coverage. The leading edge starts to align at 2.0 ML.

Figure \ref{fig3} (d) shows the TPD traces for CO$_2$ for the sub-monolayer and intermediate regimes from $\Theta_{\mathrm{i}}$ = 0.3 to 1.3 ML. The peak maximum shifts from 92.2 K at $\Theta_{\mathrm{i}}$ = 0.3 ML to 78.6 K at $\Theta_{\mathrm{i}}$ = 1.1 ML. The high temperature tail extends up to \textit{ca.} 110 K. Until a coverage of \textit{ca.} 0.8 ML the peak resembles the submonolayer peaks observed for the other molecules. For larger exposures, a sharpening of the peak maximum at 78 K indicates the commencement of multilayer formation, consistent with the appearance of zero order kinetics. This feature we use to define the CO$_2$ monolayer. The inset shows multilayer traces for coverages of $\Theta_{\mathrm{i}}$ = 1.9 to 12.2 ML. As for the other molecules, the multilayer peak shifts to higher temperatures with increasing coverage. The leading edges align at a coverage of 1.9 ML.

\subsection{Desorption energy determination}\label{Desorption energy determination}

Two different methods were used to determine the desorption energy in the sub-monolayer and multilayer coverage regimes, respectively. Multilayer desorption shows typical zero order kinetics with coincident leading edge behaviour. The desorption energy can therefore be determined by an Arrhenius analysis of the leading edge region as described previously \citep{habenschaden84, dejong90}. We obtained multilayer desorption energies of $\rm E_{des}$ = (6.7$\pm0.2)$ kJ $\rm mol^{-1}$ for CO, $\rm E_{des}$ = (11.4$\pm0.3)$ kJ $\rm mol^{-1}$ for CH$_4$, $\rm E_{des}$ = (8.3$\pm0.2)$ kJ $\rm mol^{-1}$ for O$_2$ and $\rm E_{des}$ = (21.4$\pm0.2)$ kJ $\rm mol^{-1}$ for CO$_2$. The multilayer desorption energies agree within experimental error with the values obtained from other measurements on different surfaces such as non-porous amorphous solid water ($\rm ASW{_{(np)}}$), crystalline water ice ($\rm H{_2}O{_{(c)}}$), amorphous silicate (SiO${_x}$), gold (Au), potassium bromide (KBr), graphene, amorphous silica, highly oriented pyrolytic graphite (HOPG), magnesium oxide (MgO) and methanol:water (CH$_3$OH:H$_2$O) mixed ice as listed in Table \ref{table1}. The multilayer desorption energies would be expected to be similar to the sublimation energies, also listed in Table \ref{table1} for comparison. The desorption energy of CO shows excellent agreement with sublimation energy values. However, for CH$_4$ \citep{bondi63} and O$_2$ \citep{lide93} the desorption energy is slightly higher than the sublimation energy. For CO$_2$ the measured desorption energy is significantly lower than the reported sublimation energy \citep{bryson74}, which is likely due to this being measured at a temperature of 102 K which is significantly higher than the typical desorption temperature of 80 K in the present measurements. The desorption order ($n$) and the pre-exponential factor ($\nu$) in ML s$^{-1}$ (1 ML is commonly defined as 10$^{15}$ molecules cm$^{-2}$) for multilayer desorption of CO, CH$_4$, O$_2$ and CO$_2$ from different surfaces are also listed in Table \ref{table1} for comparison. The multilayer pre-exponential factors for CH$_4$, O$_2$ and CO$_2$ are in good agreement with other studies while that for CO is several orders of magnitude smaller. We note that at a temperature of 27 K, corresponding to the onset desorption temperature for CO, transition state theory would indicate a lower limit of $k_\mathrm{B}T/h=6\times10^{11}$ ML s$^{-1}$. The reason for the low value obtained remains unclear, given the apparent agreement of the desorption energy, and we suggest use of values around $10^{12}$ ML s$^{-1}$, consistent with this limit and other works. 

For the sub-monolayer regime, the Redhead peak maximum method \citep{redhead62} is generally used to obtain the desorption energy. In the present case we, however, observe a coverage dependent desorption energy which means that there is no unique desorption energy. Rather, we have inverted the Polanyi-Wigner equation to obtain an expression for the coverage dependent desorption energy $E_{\mathrm{des}}$($\Theta$) reflecting this distribution, assuming first order desorption kinetics, \textit{i.e.,} $n=1$:

\begin{equation}
E_{\mathrm{des}}(\Theta) = -RT \ln \left[- \frac {\mathrm{d}\Theta / \mathrm{d}t}{\nu_1 \Theta}\right]
\label{equation3}
\end{equation}

This analysis was performed for each species and all initial coverages up to 1 ML, assuming a pre-exponential factor $\nu_1$ = 10$^{12}$ s$^{-1}$ for CO, CH$_4$ and O$_2$ and $\nu_1$ = 10$^{13}$ s$^{-1}$  for CO$_2$ molecule. These values are consistent with the mobile limit in transition state theory (TST) \citep{laidler40} at the typical desorption temperature range observed for each molecule. According to TST, the $\nu$ is determined by the change in entropy in going from the adsorbed to the gas phase. \citet{smith14b} used similar arguments to obtain a value of $\nu=10^{13}$ s$^{-1}$ for CO$_2$.

\begin{figure*}
\centering
    \resizebox{\hsize}{!}{\includegraphics{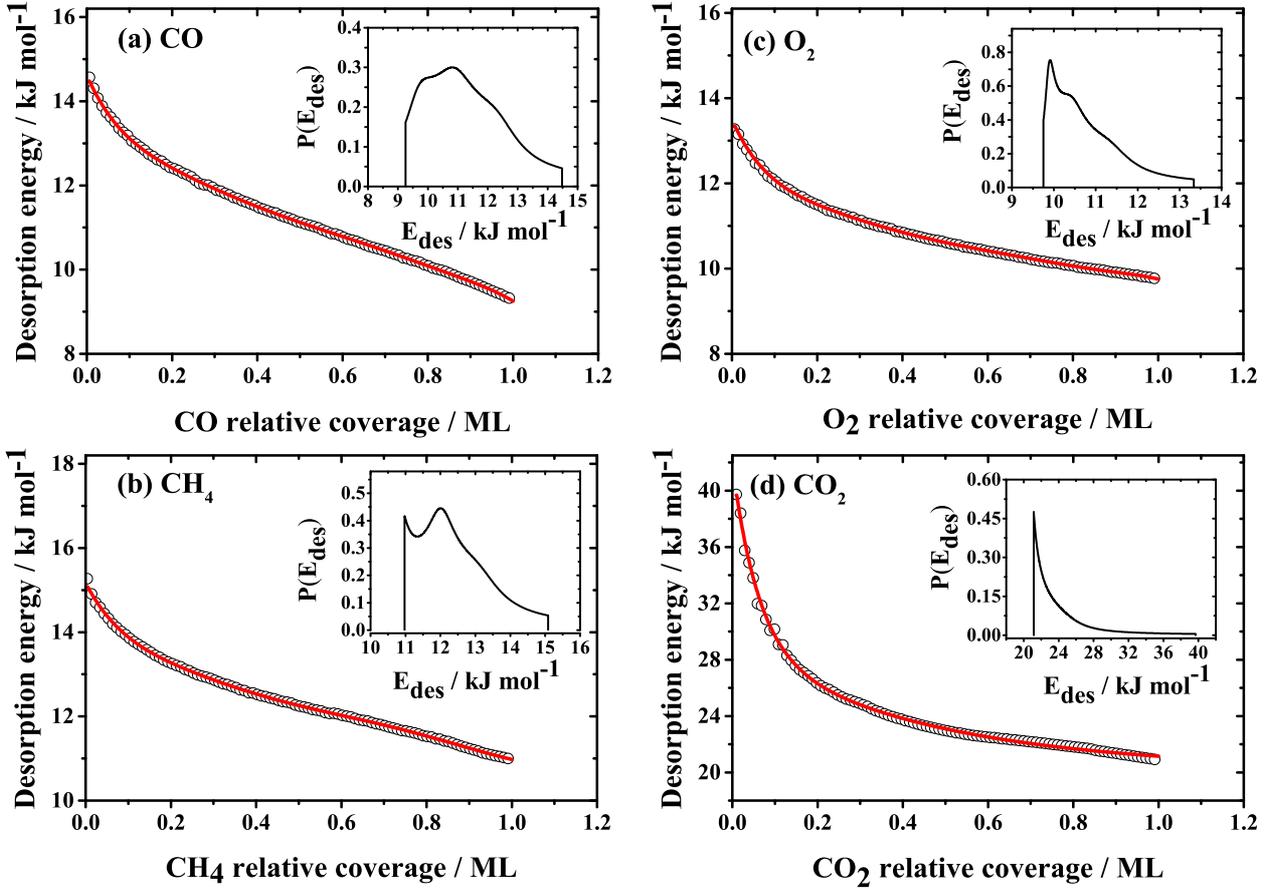}}
\caption{
Desorption energy as a function of coverage ($E_{\mathrm{des}}(\Theta)$) as obtained from the TPD traces for 1 ML of (a) CO, (b) CH$_4$, (c) O$_2$ and (d) CO$_2$ adsorbed on the forsterite(010) surface. The open circles (in grey) are the result of directly inverting the experimental traces while the solid lines (in red) are 7th order polynomial (in case of CO, CH$_4$ and O$_2$) and bi-exponential decay (for CO$_2$) fits to the data. The insets show the associated desorption energy distributions obtained using Equation \ref{equation4}.}
\label{fig4}
\end{figure*}

Figure \ref{fig4} shows the resulting desorption energy as a function of coverage obtained using Equation \ref{equation3} for $\Theta_\mathrm{i}=1$ ML for (a) CO, (b) CH$_4$, (c) O$_2$ and (d) CO$_2$. For other smaller initial coverages (data not shown), the desorption energy curves are coincident with those obtained for $\Theta_\mathrm{i}=1$ ML. Furthermore, a 7th order polynomial fit (in red) was used to fit the resulting desorption energy curve for CO, CH$_4$ and O$_2$ and a bi-exponential decay fit (also in red) for CO$_2$ as shown in Figure \ref{fig4}. It should be noted that these functional forms are purely empirical and provide a convenient way to include the coverage dependence of the desorption energy in kinetic simulations. From Figure \ref{fig4} the CO$_2$ coverage dependent desorption energy curve is clearly different to those for CO, CH$_4$ and O$_2$. In case of CO, CH$_4$ and O$_2$ the desorption energy curves extend over 4 to 5 kJ mol$^{-1}$ whereas for CO$_2$ a spread of more than \textit{ca.} 20 kJ mol$^{-1}$ is observed. The highest desorption energies are therefore 60, 29, 46 and 82 \% larger than the lowest energy sites for CO, O$_2$, CH$_4$ and CO$_2$, respectively.  

The coverage dependent desorption energy can be converted to the corresponding desorption energy distributions, $P(E_{\mathrm{des}})$ according to \citep{zubkov07}:

\begin{equation}
P(E_{\mathrm{des}}) = -\frac{\mathrm{d}\Theta}{\mathrm{d} E_{\mathrm{des}}}
\label{equation4}
\end{equation}

The resulting distributions are shown in the insets to Figure \ref{fig4}. Figure \ref{fig4} (a) demonstrates that the desorption energy of CO is dominated by values of 9.5-12.0 kJ mol$^{-1}$. In Figure \ref{fig4} (b) the desorption energy of CH$_4$ is dominated by values of 11.5-13.0 kJ mol$^{-1}$. Figure \ref{fig4} (c) illustrates that the desorption energy is dominated by values of 9.8-11.5 kJ mol$^{-1}$ for O$_2$ and Figure \ref{fig4} (d) demonstrates the desorption energy are dominated by values close to 23 kJ mol$^{-1}$ for CO$_2$. However, the desorption tail at higher temperature means that desorption energies as high as 14.5 kJ mol$^{-1}$, 15.1 kJ mol$^{-1}$, 13.4 kJ mol$^{-1}$ and 40 kJ mol$^{-1}$ are present for CO, CH$_4$, O$_2$ and CO$_2$, respectively. Table \ref{table2} lists the desorption energy ranges obtained in the present work and compares these to similar studies utilizing different substrates some of which are already listed in Table \ref{table1}. In addition, measurements on porous amorphous solid water ($\rm ASW{_{(p)}}$), meteorite nanoparticles and titanium oxide (TiO$_2$) are also listed. The desorption order ($n$) and pre-exponential factor ($\nu$) are also listed. The monolayer energy values in Table \ref{table2} correspond to the lowest desorption energy, observed at saturation coverage. For systems in Table \ref{table2} where no desorption energy range is quoted, this is due to a single desorption energy having been extracted by use of either the Redhead method \citep{mautner06, ulbricht06} or an Arrhenius analysis of the leading edge in the sub-monolayer coverage regime \citep{burke10, edridge13}. 

Table \ref{table2} furthermore shows that all four molecules exhibit a larger desorption energy distribution on forsterite(010) as compared to other surfaces in similar studies except for ASW \citep{smith16} and oxides. In the case of oxide surfaces the broader desorption energy distribution is much larger, as observed in the desorption of CH$_4$ from MgO \citep{dohnalek02} and also CO$_2$ from TiO$_2$ \citep{smith14a}. Similar to the silicate surfaces studied here, these surfaces also provide multiple adsorption sites that can result in different binding energies. For TiO$_2$(110), CO$_2$ shows site specific desorption peaks corresponding to metal and oxygen sites. In case of MgO(100), and forsterite(010) in the present work, single broad desorption peaks are instead observed, leading to coverage-dependent desorption energies. 

\begin{table*}
\caption{Kinetic parameters for the desorption of multilayer of ices along with the sublimation energy}
\label{Multilayer}
\centering
\begin{tabular}{ccccccc}
\hline
\hline
Molecule & Surface & $n$ & $\nu$ & Multilayer energy & Sublimation energy & Reference \\
& & &(ML s$^{-1}$) & (kJ $\rm mol^{-1}$) & (kJ $\rm mol^{-1}$) & \\
\hline
\multirow {11}{*}{CO} &Mg$_2$SiO$_4$(010) &0  & $6\times10^{8}$ & 6.7$\pm0.2$ &   &this work  \\
&$\rm ASW{_{(np)}}$  &0  & 7.1$\times10^{11}$  & 6.9$\pm0.2$ &   & \citet{noble12}   \\
&$\rm H{_2}O{_{(c)}}$ & 0 &7.1$\times10^{11}$  & 7.1$\pm0.5$ &   & \citet{noble12}  \\
&SiO${_x}$ & 0 &7.1$\times10^{11}$ & 6.9$\pm0.3$ &   & \citet{noble12}  \\
&Au & 0 & 7.2$\times10^{11}$ & 7.1$\pm0.1$ & & \citet{acharyya07} \\
&Polycrystalline Au &  0& 7$\times10^{11\pm1}$ & 7.1$\pm0.2$ &    & \citet{bisschop06}\\
&$\rm ASW{_{(np)}}$ &  0& 7$\pm2\times10^{11}$& 6.9$\pm0.2$ &   & \citet{collings03} \\
&KBr &  0& 6.5$^{+15.3}_{-4.6}$$\times10^{12}$ & 7.4$\pm0.3$ &  & \citet{martin14} \\
&Graphene & 0 & 4.1$\times10^{13\pm0.5}$ & 7.6$\pm0.4$ &   & \citet{smith16}  \\ 
&Amorphous silica & 0.03$\pm0.6$ & 1$\times10^{11}$&7.3$\pm0.2$ &  &  \citet{collings15} \\
&- & - & - & - & 6.9  &\citet{lide93}  \\\hline
\multirow {5}{*}{CH$_4$} &Mg$_2$SiO$_4$(010) &0  & $1.5\times10^{13}$ & 11.4$\pm0.3$ & &this work \\
& MgO(100) &0  & 1$\times10^{13\pm2}$& 13$\pm2$ &  & \citet{dohnalek02}   \\
&HOPG & 0 & 4$\times10^{15\pm1}$ & 11$\pm1$ &   & \citet{ulbricht06}  \\ 
& Graphene&0 & 2.5$\times10^{14\pm0.5}$ & 9.9$\pm0.5$ & &  \citet{smith16} \\ 
&- & - & - & - & 9.2  &\citet{bondi63}  \\\hline
\multirow {10}{*}{\oo} &Mg$_2$SiO$_4$(010) &0  & $5.4\times10^{11}$ & 8.3$\pm0.2$ &   &this work \\ 
& $\rm ASW{_{(np)}}$&0  &6.9$\times10^{11}$ &7.5$\pm0.2$ & & \citet{noble12}  \\
&$\rm H{_2}O{_{(c)}}$  &0  &7$\times10^{11}$ &7.8$\pm0.3$ &  & \citet{noble12}  \\
&SiO${_x}$ &  0&6.9$\times10^{11}$ &7.4$\pm0.3$ & & \citet{noble12} \\ 
&Au & 0 & 6.9$\times10^{11}$&7.6$\pm0.1$ &  &  \citet{acharyya07}  \\ 
&Graphene &0  & 3.2$\times10^{14\pm0.5}$&8.6$\pm0.4$ & &   \citet{smith16} \\ 
&HOPG & 0 & 8$\times10^{13\pm1}$ & 9$\pm1$ &   & \citet{ulbricht06}  \\ 
&Au&0  &7$\times10^{11\pm1}$&7.7$\pm0.2$ &  &   \citet{fuchs06} \\ 
&Amorphous silica & -0.07$\pm0.07$ & 1$\times10^{13}$&7.5$\pm0.2$ &  &  \citet{collings15} \\
&- & - & - & - & 7.3  &\citet{lide93}  \\ \hline
\multirow {12}{*}{\coo} &Mg$_2$SiO$_4$(010) &0  & $1.3\times10^{13}$ & 21.4$\pm0.2$ &   &this work \\
&$\rm ASW{_{(np)}}$ &0  &9.3$\times10^{11}$ & 18.8$\pm0.6$ &  & \citet{noble12}  \\
&$\rm H{_2}O{_{(c)}}$  &0  & 9.5$\times10^{11}$& 19.6$\pm0.7$ &  &  \citet{noble12} \\
&SiO${_x}$ &0  &9.3$\times10^{11}$ & 18.9$\pm0.7$ &  & \citet{noble12}  \\ 
&KBr & 0 & 5.2$^{+4.3}_{-1.6}$$\times10^{12}$& 21.7$\pm0.2$ & & \citet{martin14}  \\ 
&\coo &0  && 22.4$\pm0.4$ &  & \citet{sandford90}  \\ 
&HOPG &0  &1.1$\pm0.1\times10^{11}$ & 24.8$\pm1.6$ &  & \citet{burke10}  \\ 
&$\rm ASW{_{(np)}}$ &0 &2.1$\times10^{11}$  & 25.4$\pm1.4$ & &  \citet{edridge13} \\ 
& CH$_3$OH:H$_2$O &0 &2.3$\times10^{10}$  & 24.3$\pm2.4$ & &  \citet{edridge13} \\ 
&$\rm H{_2}O$ &0  && 23.8$\pm1.7$ &  &  \citet{sandford90} \\  
&HOPG &0  &6$\times10^{14\pm1}$ & 23$\pm2$ &  &  \citet{ulbricht06} \\  
&- & - & - & - & 27.2$\pm0.4$  &\citet{bryson74}  \\\hline
\hline
\label{table1}
\end{tabular}
\end{table*}

\begin{table*}
\centering
\caption{Kinetic parameters for the desorption of monolayer and sub-monolayer}
\label{Monolayer}
\begin{tabular}{ccccccc}
\hline
\hline
Molecule & Surface & $n$ & $\nu$ & Monolayer energy  & Energy distribution & Reference \\
& & &(s$^{-1}$) & (kJ $\rm mol^{-1}$) & (kJ $\rm mol^{-1}$) & \\
\hline
\multirow {12}{*}{CO} &Mg$_2$SiO$_4$(010) & 1 & 1$\times10^{12}$ & 9.3$\pm0.1$ & 14.5-9.2  & this work  \\
&$\rm ASW_{(p)}$  & 1 & 1$\times10^{12}$  & & 14.2-8.6  & \citet{he16}   \\
&$\rm ASW_{(np)}$ & 1 &1$\times10^{12}$  &  & 12.3-8.1  & \citet{he16}  \\
&$\rm ASW{_{(np)}}$  &1  & 1$\times10^{12}$  & 7.2 &10.9-7.2   & \citet{noble12}   \\
&$\rm H{_2}O{_{(c)}}$ & 1 &1$\times10^{12}$  & 8.4 & 11.1-8.4  & \citet{noble12}  \\
&SiO${_x}$ & 1 &1$\times10^{12}$ & 7.2&11.8-7.2   & \citet{noble12}  \\
&Graphene & 0 &1.9$\times10^{13}$ & 12.5$\pm0.6$ & 12.5-7.5  & \citet{smith16}  \\
&$\rm ASW{_{(np)}}$ & 1 & 3.5$\times10^{16}$ & 11.8$\pm0.6$ & 16-11  & \citet{smith16} \\
&HOPG & 1 & 2$\times10^{14\pm1}$&13$\pm1$ &  &  \citet{ulbricht06}  \\ 
&Meteorite & 1 & 1$\times10^{13}$&13.5$\pm3$ &  &  \citet{mautner06}  \\ 
&$\rm ASW_{(np)}$ & 1 & 5$\pm1$$\times10^{14}$&9.8$\pm0.2$ &  &  \citet{collings03} \\
&Amorphous silica & 1 & 1$\times10^{11}$& & 12.2-8.2 &  \citet{collings15} \\ \hline
\multirow {6}{*}{CH$_4$} &Mg$_2$SiO$_4$(010) &1  & 1$\times10^{12}$ & 11$\pm0.1$ & 15.1-11.5 &this work \\
& $\rm ASW_{(np)}$ & 1 & 1$\times10^{12}$& & 12.2-9.3 & \citet{he16}   \\
& Graphene  & 0 & 2.1$\times10^{13}$& 14.9$\pm0.7$ & 15-10 &  \citet{smith16} \\
&$\rm ASW{_{(np)}}$ & 1 &9.8$\times10^{14}$ & 11.4$\pm0.6$ & 16-11.4 & \citet{smith16}  \\ 
&MgO(100) & 1 &1$\times10^{13}$ &  & 20-11& \citet{dohnalek02}  \\ 
&HOPG&1  &4$\times10^{15\pm1}$ &17$\pm1$  & & \citet{ulbricht06}  \\  \hline
\multirow {9}{*}{\oo} &Mg$_2$SiO$_4$(010) & 1 & 1$\times10^{12}$ & 9.8$\pm0.1$ & 13.4-9.7  &this work \\
&$\rm ASW{_{(np)}}$  &1  & 1$\times10^{12}$  &7.6 &9.7-7.6   & \citet{noble12}   \\
&$\rm H{_2}O{_{(c)}}$ & 1 &1$\times10^{12}$  & 8.1 &9.6-8.1   & \citet{noble12}  \\
&SiO${_x}$ & 1 &1$\times10^{12}$ & 7.7 &10.4-7.7   & \citet{noble12}  \\
& $\rm ASW_{(np)}$& 1 &1$\times10^{12}$ & & 11.1-7.7& \citet{he16}  \\
&Graphene & 0 &1.1$\times10^{13}$ &11.8$\pm0.6$& 12-8.5 & \citet{smith16}  \\
&$\rm ASW{_{(np)}}$ &1  &5.4$\times10^{14}$ &9.2$\pm0.5$ & 14-9.2  & \citet{smith16} \\ 
&HOPG & 1 & 8$\times10^{13\pm1}$&12$\pm1$ &  &  \citet{ulbricht06}  \\ 
&Amorphous silica & 1 & 1$\times10^{13}$& & 11.9-8 &  \citet{collings15} \\ \hline
\multirow {10}{*}{\coo} &Mg$_2$SiO$_4$(010) & 1 & 1$\times10^{13}$ & 20.9$\pm0.3$ & 40- 21.1 &this work \\
&Olivine &1 & 1$\times10^{19}$ &  &  80-30&  \citet{smith14b} \\
&$\rm ASW{_{(np)}}$  &1  & 1$\times10^{12}$  & 18.6 &19.5-18.6   & \citet{noble12}   \\
&$\rm H{_2}O{_{(c)}}$ & 1 &1$\times10^{12}$  & 19.6 &20.9-19.6   & \citet{noble12}  \\
&SiO${_x}$ & 1 &1$\times10^{12}$ & 18.9 & 25-18.9  & \citet{noble12}  \\
& TiO$_2$ & 1&  4.6$\times10^{16}$ &  &  64-36&  \citet{smith14a} \\ 
& HOPG &1 &6$\times10^{14\pm1}$  & 24$\pm2$ & &  \citet{ulbricht06} \\ 
&HOPG&0.73$\pm0.02$  & 5.3$\pm0.8\times10^{14}$& 19.8$\pm2.9$ &   & \citet{burke10}  \\ 
& HOPG &0.73$\pm0.2$ &9.9$\times10^{14\pm0.9}$  & 20.2$\pm2.3$ & &  \citet{edridge13} \\ 
&$\rm ASW{_{(np)}}$ &1.09$\pm0.06$ &2.7$\times10^{7\pm2}$  & 17.5$\pm7.5$ & &  \citet{edridge13} \\ 
\hline
\label{table2}
\end{tabular}
\end{table*}

\section{Discussion}

Experimental TPD traces for all four molecules described here show multilayer desorption sharing a common leading edge at high initial coverages. Multilayer desorption occurs from several layers of molecules on the surface and hence the underlying surface does not influence the desorption kinetics as strongly as for the monolayer, as adsorbate-adsorbate interactions dominate over adsorbate-substrate interactions. This is evident from the multilayer desorption energy of $E_{\mathrm{des}}=6.7\pm0.2$ kJ mol$^{-1}$ for CO ice on forsterite(010) which is in good agreement with previous studies of CO deposited on a variety of substrates as listed in Table \ref{table1}. The agreement in multilayer desorption energy value is excellent in the case of CO desorption from amorphous silica \citep{collings15} and from amorphous silicates \citep{noble12, collings15}. In the case of CH$_4$, there are no multilayer desorption measurements from silicate surfaces. The present value is, however, in good agreement with the multilayer desorption energy for desorption from HOPG \citep{ulbricht06}.The multilayer desorption energy of O$_2$ agrees with those obtained for graphene \citep{smith16} and for amorphous silicate and silica surfaces \citep{noble12, collings15}. For CO$_2$ the multilayer desorption energy of $E_{\mathrm{des}}=21.4\pm0.2$ kJ mol$^{-1}$ shows reasonable agreement with values reported previously as listed in Table \ref{table1}. It is evident that there is some spread in the multilayer desorption energies obtained for CO$_2$. Our multilayer desorption energy is in excellent agreement with the values reported for a KBr substrate by \citet{martin14}. However, the values reported for an HOPG surface \citep{ulbricht06, burke10} and by \citet{edridge13} for ASW and a methanol-water mixed ice surface, is somewhat higher. On the other hand, the values reported by \citet{noble12} for water and silicate are somewhat lower than our value.

For sub-monolayer coverages we see a trailing edge for all molecules which extends to higher temperatures than would be expected for simple first order desorption from a single type of adsorption site. Such behaviour is similar to our previous studies of NH$_3$ desorption from forsterite(010) \citep{suhasaria15}. We previously concluded that this desorption tail arises from the presence of a variety of adsorption sites on the forsterite surface due to presence of species such as O, Mg, Si and Fe which is also consistent with previous work by \citet{smith14b} on crystalline olivine(011). \citet{thrower09} and \citet{collings15} showed that the amorphous silica surface also presents a range of binding sites which influences the desorption kinetics of C$_6$H$_6$ and volatile species such as O$_2$ and CO. Similarly, NH$_3$ desorption showed a similar behaviour when deposited on an amorphous silicate surface \citep{he15}. Furthermore, we observe that trailing edges for different initial coverages are aligned which indicates that the molecules are able to diffuse on the surface and find the high energy binding sites prior to adsorption. However, the nature of the TPD experiment means that we are unable to determine at which point this mobility becomes possible \textit{i.e.} the molecules may begin to diffuse at any point during the heating process. Thus, the coverage dependent desorption energies we extract represent the available sites rather than necessarily providing insight into the population of these sites upon adsorption.

We observe that for CO, O$_2$ and CO$_2$ on the forsterite(010) surface, generally higher desorption energies are present compared to previous measurements on amorphous silicate and silica surfaces \citep{noble12, collings15}, as well as on H$_2$O or HOPG, using similar pre-exponential factors as evident from Table \ref{table2}. Overall it is apparent that these molecules tend to bind with higher energy to the crystalline silicate forsterite(010) surface. For CH$_4$, there are no reported desorption measurements for silicate surfaces. The desorption of CO$_2$ from the olivine surface has been measured by \citet{smith14b}, yielding a desorption energy range of 30-80 kJ mol$^{-1}$. These energies are significantly higher than those for forsterite(010) which are in the range 21.1-40 kJ mol$^{-1}$. This could be due to the use of a pre-exponential factor seven orders of magnitude larger than that employed in the present study. Such a high value for pre-exponential is difficult to reconcile with the desorption of a small molecule such as CO$_2$ where significant entropic effects are not likely to play a large role in increasing the value several orders of magnitude from the mobile limit.

The observed appearance of the multilayer peak prior to completion of the monolayer for CO, CH$_4$ and O$_2$ is indicative of island growth on the surface. At low coverages, these molecules are bound to the silicate surface due to the strong adsorbate-substrate interaction revealed by the higher desorption energy tail. However, as the coverage increases progressively lower energy binding sites are populated and the probability of an incoming molecule landing on top of adsorbate covered regions becomes more likely. The clear separation between the monolayer and multilayer desorption features results from the multilayer desorption energy being somewhat smaller than the lowest monolayer desorption energy.  For CO$_2$ there appears to be be a gradual progression from occupation of monolayer sites to formation of the monolayer, most likely arising from the overlap of the desorption energies from the most weakly binding monolayer sites and the multilayer. Similar behaviour was observed in the case of desorption of  NH$_3$ from the same forsterite(010) surface \citep{suhasaria15}.

\begin{figure}
\centering
    \resizebox{\hsize}{!}{\includegraphics{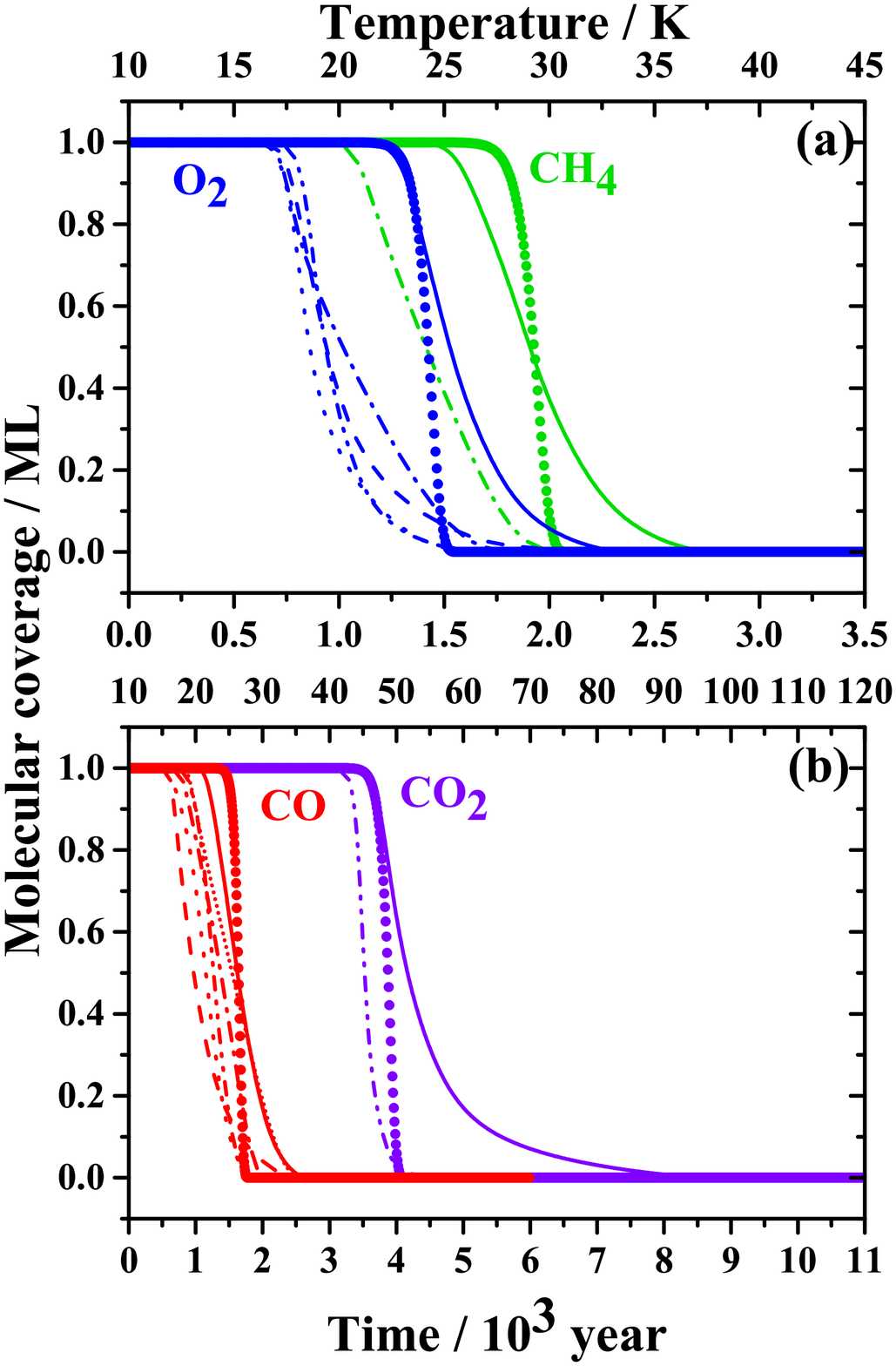}}
\caption{Simulated coverage remaining on the surface plotted against desorption time for 1 ML of (a) O$_2$ (blue) and CH$_4$ (green) and (b) CO (red) and CO$_2$ (violet) at a heating rate of 1 K century$^{-1}$. Surfaces used for comparison are: $\rm ASW_{(np)}$ (dash dot lines; \citep{he16}), $\rm ASW_{(p)}$(short dot lines; \citep{he16}), $\rm ASW_{(np)}$ (dot lines; \citep{noble12}), H${_2}$O${_{(\mathrm{c})}}$ (dash dot dot lines; \citep{noble12}), SiO$_x$ (dash lines; \citep{noble12}), forsterite(010) (solid lines, this work). The simulated desorption coverage curves for constant desorption energy are shown with closed circles}
\label{fig5}	
\end{figure}

\section{Astrophysical Implications}
Laboratory experiments can provide the necessary kinetic parameters to simulate astrophysical processes. However, laboratory experiments are performed on a much shorter time scale compared to the typical lifetime of interstellar clouds (\textit{ca.} $10^6$ yr.). During this timescale, in dense clouds, icy mantles are formed on the grain surfaces through a combination of freeze-out (\textit{e.g.} CO) and recombinative surface reactions (\textit{e.g.} to form H$_2$O and CH$_3$OH). As these clouds collapse, eventually leading to star formation, the temperature of the grains rises and molecules thermally desorb. To understand thermal desorption in the early stages of star formation on an astrophysical time scale, we have used a heating rate of 1 K century$^{-1}$as proposed by \citet{viti99} for hot cores. We have used our experimentally derived kinetic parameters to simulate the desorption traces for 1 ML coverage by numerically integrating the Polanyi-Wigner equation and incorporating the coverage dependent desorption energy ($E_{\mathrm{des}}(\Theta)$) determined from our laboratory measurements. First order desorption kinetics and a pre-exponential factor of $\rm 10^{12}$ or $\rm 10^{13} s^{-1}$ depending on the molecules, as discussed above, were used in performing the simulations. A plot of the surface coverage against time was obtained. We have additionally performed similar simulations for water ice \citep{noble12, he16} and silicate \citep{noble12} surfaces using the kinetic parameters reported in the other studies listed in Table \ref{table2}. In Figure \ref{fig5}, we compare the simulations for the different substrates. We note that an apparent mistake in one of the coefficients for $E_\mathrm{des}(\Theta)$ precludes comparison with CO$_2$ on the amorphous silicate surface reported by \citet{noble12}.

Figure \ref{fig5} shows the surface coverage versus desorption time for CO (red), CH$_4$ (green), O$_2$ (blue) and CO$_2$ (violet). The substrate used is indicated by the line style, $\rm ASW_{(np)}$ (dash dot lines; \citep{he16}), $\rm ASW_{(p)}$(short dot lines; \citep{he16}), $\rm ASW_{(np)}$ (dot lines; \citep{noble12}), H${_2}$O${_{(\mathrm{c})}}$ (dash dot dot lines; \citep{noble12}), SiO$_x$ (dash lines; \citep{noble12}), forsterite(010) (solid lines; this work).

In addition, the simulated desorption coverage curves for a constant desorption energy are shown in closed circles for each molecule to compare the effect of coverage dependent desorption energy. The constant desorption energy is calculated from the Redhead approximation method as mentioned in Section \ref{Desorption energy determination}. For this calculation the maximum peak temperature is chosen at the 1 ML coverage which gives values of 10.8 kJ mol$^{-1}$ (CO), 12 kJ mol$^{-1}$ (CH$_4$), 9.9 kJ mol$^{-1}$ (O$_2$) and 21.1 kJ mol$^{-1}$ (CO$_2$), representing the lowest energy sites present in the distribution. It is evident from the Figure 5 (a), that the onset of desorption of both CH$_4$ and O$_2$ occurs at a higher temperature and a later time (1200 and 1450 yr.) from the forsterite(010) as compared to the other surfaces and these molecules remain longer on the crystalline forsterite surface before they are fully desorbed. For CH$_4$ desorbing from the forsterite(010) surface (green solid lines) takes 750 years longer to fully desorb in comparison to the $\rm ASW_{(np)}$ (green dash dot lines) and O$_2$ from forsterite(010) surface takes 500 years more on average compared to other surfaces shown in Figure \ref{fig5} (a).

Similarly, from Figure \ref{fig5} (b), we observe that much of the CO$_2$ remains on the forsterite(010) surface (violet solid lines) for 3900 years longer in comparison to the $\rm H{_2}O{_{(c)}}$ surface (violet dash dot dot lines) and the onset of desorption is 300 years later on forsterite(010) surface compared to $\rm H{_2}O{_{(c)}}$ surface. CO also shows a delay in the onset of desorption from forsterite(010) surface by an average of 200 years as evident from Figure \ref{fig5} (b) and remains on the forsterite surface for longer times. We observe that the desorption times for both O$_2$ and CO from crystalline silicate (forsterite) are longer than from amorphous silicates by \textit{ca}. 250 and 550 years respectively.

Longer desorption timescales mean that more molecules remain in the solid state for a longer time, increasing the probability that they actively take part in grain surface chemistry, compared to current astrochemical gas-grain models in which a single desorption energy is included.

\section{Conclusions}
We have investigated the thermal desorption of CO, CH$_4$, O$_2$ and CO$_2$ from a forsterite(010) model grain surface representative of the silicate family of grains. We have characterized this stable crystalline silicate surface using AFM measurements which confirm atomic flatness over large areas. Morphological effects as observed in the case of rough and amorphous surfaces are therefore expected to be less important with the adsorption and desorption behaviour depending on the chemical nature of the sites available. Thermal desorption in the multilayer coverage regime for all the species investigated display coincident leading edges consistent with zero order desorption kinetics. This behaviour is in agreement with previous studies and demonstrates that, beyond the first few layers, the desorption behaviour is effectively insensitive to the nature of the underlying substrate. On the other hand, in the sub-monolayer coverage regime, all four molecules display coincident trailing edges that extend to high temperature. This indicates the presence of a coverage dependent desorption energy which could arise, either as a result of the present of a distribution of sites on the surface which present different binding energies, or the effect of increasing inter-molecular repulsion between adsorbate molecules with increasing coverage. While our measurements allow us to determine the range of possible desorption energy values, the TPD technique employed does not allow us to distinguish between these two possibilities. We, however note, that a variety of adsorption sites on the crystalline forsterite(010) surface is not unreasonable, and the observed behaviour may arise as a result of a combination of the two effects. We observe that for CO, CH$_4$ and O$_2$, the growth of multilayers commences prior to saturation of the monolayer peak while, CO$_2$, shows a desorption behaviour which smoothly shifts from the sub-monolayer to multilayer as a result of the similarity between the multilayer and minimum monolayer desorption energies. In all cases, the forsterite(010) surface results in desorption energy distributions that extend to higher energy than other surfaces such as ASW, crystalline water and amorphous silica, suggesting the presence of higher energy binding sites. By using kinetic simulations employing astrophysically relevant heating rates, the onset of desorption for these species thus occurs at later times for forsterite(010) than other surfaces, and the extended desorption timescale results in significantly longer residence times for the most strongly bound molecules when compared to the other substrates.

\section{Acknowledgements}
We acknowledge financial support from the European Commission's 7th Framework Programme through the "LASSIE" ITN under Grant Agreement Number 238258, the NRW International Graduate School of Chemistry, M{\"u}nster. We would also like to acknowledge the European union (EU) and Horizon 2020 funding awarded under the Marie Sklodowska-Curie action to the EUROPAH consortium, grant number 722346. 












\bibliographystyle{mnras}
\bibliography{References} 

\bsp	
\label{lastpage}
\end{document}